\documentclass[epj,final]{svjour}
\usepackage{graphicx}
\usepackage{xspace}
\newcommand{\rcm}{\mbox{cm$^{-1}$}\xspace}
\newcommand{\Xstate}{X$^1\Sigma^+$\xspace}
\newcommand{\Bstate}{B$^1\Pi$\xspace}
\newcommand{\astate}{a$^3\Sigma^+$\xspace}
\usepackage{amsmath,amssymb}

\begin{document}
\title{Population redistribution in optically trapped polar molecules}

\author{J. Deiglmayr$^{1,}$\thanks{Present address: ETH Z\"urich, Laboratory of Physical Chemistry, Wolfgang-Pauli-Str. 10, CH-8093 Zurich} \and
M. Repp$^1$ \and
O. Dulieu$^2$ \and
R. Wester$^3$ \and
M. Weidem\"uller$^{1,}$\thanks{e-Mail: weidemueller@physi.uni-heidelberg.de}
}

\institute{$^1$Ruprecht-Karls-Universit\"at Heidelberg, Physikalisches
Institut, Philosophenweg 12, 69120 Heidelberg, Germany, \\
$^2$Laboratoire Aim\'e Cotton, CNRS, Universit\'e Paris Sud XI, Orsay, France, \\
$^3$Universit\"at Innsbruck, Institut f. Ionenphysik und Angewandte Physik, Technikerstraße 25/3, A-6020 Innsbruck
}

\date{Received: date / Revised version: date}

\abstract{We investigate the rovibrational population redistribution of polar molecules in the electronic ground state induced by spontaneous emission and blackbody radiation. As a model system we use optically trapped LiCs molecules formed by photoassociation in an ultracold two-species gas. The population dynamics of vibrational and rotational states is modeled using an \emph{ab-initio} electric dipole moment function and experimental potential energy curves. Comparison with the evolution of the $v''$=3 electronic ground state yields good qualitative agreement. The analysis provides important input to assess applications of ultracold LiCs molecules in quantum simulation and ultracold chemistry.}

\maketitle
%
% introduction
  Ultracold gases of dipolar molecules have been proposed as candidates for the exploration of quantum phases in dipolar gases~\cite{Pupillo2008b}, the development of quantum computation techniques~\cite{DeMille2002}, or precision measurements of fundamental constants~\cite{zelevinsky2008}. Recently tremendous progress has been made in the realization of cold and ultracold gases made out of polar molecules \cite{Bethlem2000,Rieger2005,Ni2008,Deiglmayr2008b}. The presence of an electric dipole moment (EDM) in the electronic ground state leads to a coupling of the internal molecular state to the thermal environment via black-body radiation (BBR). Thus if the system is prepared in a certain internal state, this state will be altered after a finite time. In particular this yields a limited lifetime even for molecules prepared in the absolute rovibrational ground state. The thermalization of cold OH and OD molecules with the environment was observed experimentally~\cite{Hoekstra2007} and theoretical studies were performed for a number of polar species~\cite{Vanhaecke2007,Buhmann2008}. For the molecular ions HD$^+$ or MgH$^+$ rotational cooling relying upon fast BBR-driven rotational relaxation was recently demonstrated by \cite{Schneider2010} and \cite{Staanum2010}, respectively. As a second consequence of the existence of an EDM, spontaneous decay of electronic ground state molecules into more deeply bound rovibrational levels is possible. For the molecular ion HD$^+$ radiative decay leads to vibrational relaxation into $v"=0$ in less than 1~s~\cite{Amitay1994,Amitay1998}. Radiative transitions rates for polar alkali dimers have been calculated~\cite{Zemke2004,Mayle2007}, but were generally considered to be too small to be of experimental relevance.

  Here we model the population redistribution in LiCs in the electronic ground state induced by spontaneous emission and BBR. We compare this simulation to measurements on an  ultracold gas of LiCs molecules in the electronic ground state, confined in an optical dipole trap insensitive to the internal state of the molecules. An ultracold gas of LiCs, the most polar alkali dimer~\cite{Aymar2005}, in deeply bound levels of the \Xstate state can be efficiently formed by photoassociation ~\cite{Deiglmayr2008b,Deiglmayr2009a}. We have measured the EDM of deeply bound ground state levels to be as high as 5.5\,Debye~\cite{Deiglmayr2010}, making this molecule a promising candidate for the realization of an ultracold gas with strong dipolar interactions. Here we observe significant dipole transition rates between vibrational ground state levels, enhanced by the strong dependence of the dipole moment function on the internuclear distance which is associated with its large average value.

\section{\textit{Ab-initio} modeling of transition rates}

  A transition between levels $i$ and $j$ with wave functions $\Psi_i$ and $\Psi_j$ is proportional to the dipole matrix element $\mu_{i,j}=\langle \Psi_i| \mu(R) | \Psi_j \rangle$, where $\mu(R)$ is the EDM function. For vibrational transitions only the variation of $\mu$ with the internuclear distance $R$ leads to a finite transition dipole moment. The EDM function was calculated \textit{ab-initio} and published previously~\cite{Aymar2005}. Rovibrational wave functions and overlap integrals are calculated with Level 8.0~\cite{Leroy2007} using a precise experimental potential energy curve~\cite{Staanum2007}. The resulting EDMs of low lying vibrational levels are in very good agreement with measured values~\cite{Deiglmayr2010}. For modeling the spontaneous decay we calculate the Einstein A coefficients
  \begin{eqnarray*}
  A_{ij}\equiv A_{v'J',v''J''}= \frac{8 \pi^2 S_{J',J''}}{3 \varepsilon_0 \hbar \,(2J'+1)}\tilde{\nu}_{v'J',v''J''}^3 \mu_{v'J',v''J''}^2
  \end{eqnarray*}
  with the transition dipole moment $\mu_{v'J',v''J''}$ (in SI units), a transition energy $\tilde{\nu}_{v'J',v''J''}$ (in wave numbers), and the H\"{o}nl-London (HNL) Factor $S_{J',J''}$ as defined in Ref.~\cite{hansson2005}.  The coupling to BBR from the surrounding walls of the vacuum chamber is given by the Einstein B coefficients for stimulated absorption and emission multiplied by the spectral density of the BBR from an environment with given temperature T as
  \begin{eqnarray*}
  \tilde{B}_{ij} \equiv  \tilde{B}_{v'J',v''J''} =
  A_{v'J',v''J''}\frac{1}{e^{h \tilde{\nu} c/k T}-1}  \hspace{0.5cm}.
  \end{eqnarray*}

  \begin{table*}
    \begin{center}
      \resizebox{0.6\textwidth}{!}{
      \begin{tabular}{|c|c|c|c|c|c|c|c|c|c|}\hline
      $v''$     & 0 & 1 & 2   & 3   & 4   & 5   & 6 & 7   & 8 \\ \hline
      Pop. [\%] & 0.0 & 0.0 & 0.5 & 2.4 & 5.1 & 2.9 & 0.0 & 2.5 & 0.3 \\ \hline
      $A_{v''+1,v''}$  [10$^{-3}$/s] & 22.7 & 41.8 & 57.4 & 69.6 & 78.6 & 84.5 & 87.5 & 87.9 & 85.8  \\ \hline
      $\tilde{B}_{v''+1,v''}$ [10$^{-3}$/s] &  15.7 & 29.3 & 41. & 50.6 & 58.1 & 63.6 & 67.1 & 68.6 & 68.3 \\ \hline
      $E_{v''}$ [cm$^{-1}$] & 0.0 & 182.7 & 363.4 & 542.0 & 718.6 & 893.2 & 1065.7 & 1236.1 & 1404.4 \\ \hline     \end{tabular}}
      \vspace{0.5cm}
      \caption{Relative population strength of the relevant lowest vibrational levels of the \Xstate state after spontaneous decay from \Bstate,$v'$=26; total vibrational rates $A_{v',v''}$ and $\tilde{B}_{v',v''}$ (for T~=~293~K), summed over rotational quantum numbers in $A_{v'J',v''J''}$ and $\tilde{B}_{v'J',v''J''}$; energy of vibrational levels $E_{v''}$. Rates for transitions with $\vartriangle$$v''\geq2$ are significantly smaller and are thus not explicitly given, but are included in the simulation.}
      \label{tab:relpop}
    \end{center}
  \end{table*}

  \begin{table*}
    \begin{center}
      \resizebox{0.6\textwidth}{!}{
      \begin{tabular}{|c|c|c|c|c|c|c|c|c|c|}\hline
      $J''$     & 0 & 1 & 2   & 3   & 4   & 5   & 6 & 7   & 8 \\ \hline
      $A_{0,J''+1,0,J''}$  [10$^{-6}$/s] & 0.2 & 1.6 & 5.9 & 14.6 & 29.2 & 51.2 & 82.2 & 123.8 & 177.3 \\ \hline
      $\tilde{B}_{0,J''+1,0,J''}$ [10$^{-3}$/s] &  0.3 & 0.7 & 1.5 & 2.5 & 3.9 & 5.5 & 7.3 & 9.5 & 11.9 \\ \hline
      \end{tabular}}
      \vspace{0.5cm}
      \caption{Rate constants $A_{v',J',v'',J''}$ and $\tilde{B}_{v',J',v'',J''}$ (for T~=~293~K), for pure rotational transitions in the vibrational ground state. The rotational constant for $v''$=0 is 0.1874~cm$^{-1}$.}
      \label{tab:rotrates}
    \end{center}
  \end{table*}

  In Tab.~\ref{tab:relpop} the resulting rates at T~=~293~K for the lowest vibrational levels are given. The given value for $\tilde{B}_{v'=1,v''=0}$$\equiv$$\tilde{B}_{v'=0,v''=1}$ corresponds to a depopulation time constant of 65.4\,s for molecules in the absolute ground state, in good agreement with previously calculated values~\cite{Vanhaecke2007,Buhmann2008}. It is noteworthy that at cryogenic temperatures (77\,K) this depopulation time constant is lengthened to 1.3$\times$10$^3$~s.  Already at room temperature rotational heating due to BBR is greatly suppressed by the low spectral density at the relevant transition frequencies. Only for $J''>$12 rotational heating rates become comparable to experimental loss rates from the trap, which are on the order of tens of seconds. Rates for rotational transitions in the vibrational ground state are listed in Tab.~\ref{tab:rotrates}.

  The evolution of an initial distribution of populated ground state levels is finally given by the following rate-equation for the population $N_i$ in a level $i$ (the single index $i$ enumerates rovibrational levels $v''$,$J''$ in energetic ordering)
  %\begin{widetext}
  %\begin{equation}
  \begin{eqnarray}
    \dot{N}_i(t) =  \sum_{j>i} A_{ji} N_j(t) - \sum_{j<i} A_{ij}N_i(t) + \nonumber & \\
     + \sum_{j\neq i} \tilde{B}_{ji}N_j(t) - \sum_{j\neq i} \tilde{B}_{ij}N_i(t) - \Gamma N_i(t) &
    \label{eq:rateeq}
  \end{eqnarray}
  %\end{equation}
  %\end{widetext}
  where $\Gamma$ is a decay constant accounting for losses from the trap. The value for $\Gamma$=21s is taken from measurements (see below) and is assumed to be independent of $v''$.

\begin{figure}
   \begin{center}
   \includegraphics[width=\columnwidth,clip]{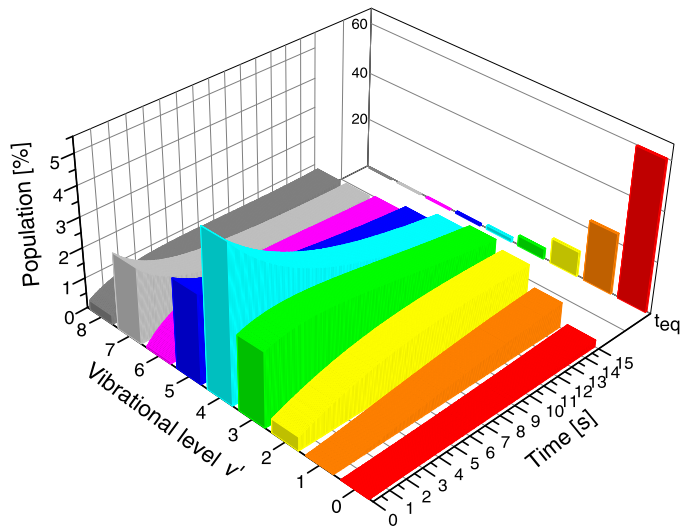}
   \caption{Solution of Eq.~\ref{eq:rateeq} for the population in the lowest nine vibrational levels of the \Xstate after spontaneous decay from \Bstate,$v'$=26,$J'$=1. The population is given as percentage of the initial total ground state population. Shown is also the population (rescaled to unity) after t$_{eq}$$\equiv$$5\times A_{1,0}^{-1}\sim200$~s.}
   \label{fig:timeevolutionAll}
   \end{center}
  \end{figure}

\begin{figure}
   \begin{center}
   \includegraphics[width=0.85\columnwidth,clip]{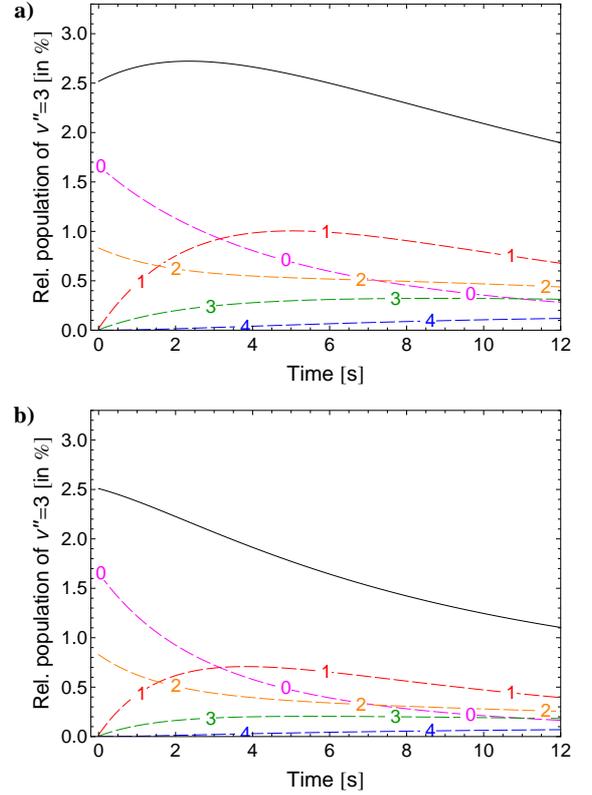}
   \caption{Simulated time evolution of the population in \Xstate,$v''$=3. Shown are the total signal (solid line) and the population of the rotational sublevels (dashed lines, labeled by $J''$). a) Simulation based on a first order loss mechanism (eq.~\ref{eq:rateeq}), b) simulation based on a second order loss mechanism (eq.~\ref{eq:rateeq2b}).}
   \label{fig:timeevolution3}
   \end{center}
  \end{figure}

  In the formation of ultracold molecules by photoassociation (PA)~\cite{thorsheim1987,jones2006}, the initial population of ground state levels is determined by the Einstein A coefficients for the spontaneous decay from the excited PA state. For the \Bstate state of LiCs we have calculated this initial population for different PA levels and found agreement with experimental observations~\cite{Deiglmayr2009a}. The Frank-Condon factors between a given level in the \Bstate state and levels in the \Xstate state show two maxima corresponding to a maximized overlap at the inner respectively the outer classical turning point of the vibrational motion. This leads to a typical bimodal distribution of the populated ground state levels with significant populations in deeply bound levels below $v''\lesssim8$ and in a small band of higher lying vibrational levels which shifts with the employed photoassociation resonance. The ground state population after PA via \Bstate,$v'$=26,$J'$=1 is given in Tab.~\ref{tab:relpop} for the deepest bound vibrational levels. The initial rotational distribution is determined by the HNL factors for the decay from the negative parity component of the excited PA level (due to the low collision temperature of the atoms in the experiment $s$-wave scattering dominates~\cite{Deiglmayr2008b}), as verified experimentally by depletion spectroscopy. Starting from this initial distribution the rate equation Eq.~(\ref{eq:rateeq}) is integrated numerically in order to derive the time evolution of the population, where convergence of the results is reached for time steps smaller than 50~ms. As can be seen in Fig.~\ref{fig:timeevolutionAll} the initial population distribution in the lowest vibrational levels of the \Xstate state is strongly modified. Most levels show either a steady increase or decrease essentially due to vibrational redistribution with adjacent states. As an exception the population in the $v''$=3 level exhibits a clear maximum after roughly 3~s, presenting a concise feature for testing the validity of the model. The time evolution of the population in this level is shown in more detail in  Fig.~\ref{fig:timeevolution3}~a) together with the individual contributions from the rotational sublevels. The significant increase of the population in $v''$=3,$J''$=1, being caused by the decay of molecules from $v''$=4 and the cycling of population between neighboring levels due to BBR with comparable rates, indicates stronger dynamics of the process than visible from the vibrational populations only.

  We test the sensitivity of the model to the employed \textit{ab-initio} dipole moment function by repeating the calculation using a different dipole moment function published in Ref.~\cite{Sorensen2009}. The outcome of the simulation is modified only slightly and shows the same qualitative time evolution for the total population in vibrational levels $v''\le 8$.

  As an alternative test we replace the first order loss term $\Gamma N_i(t)$ in Eq.~\ref{eq:rateeq} by a second order loss term where the values for the rate constant $\beta_{\textrm{LiCs-LiCs}}$=2$\times10^{-10}$~cm$^{3}/s$, the initial mean molecular density $\overline{n}_{0,\textrm{LiCs}}$=2$\times10^{8}$~cm$^{-3}$, and the initial total number of molecules $N_\textrm{0,\textrm{tot}}$=5$\times10^{3}$ are taken from experimental estimates (see below):

  \begin{eqnarray}
    \dot{N}_i(t) =  \sum_{j>i} A_{ji} N_j(t) - \sum_{j<i} A_{ij}N_i(t) + \nonumber \label{eq:rateeq2b} & \\
     + \sum_{j\neq i} \tilde{B}_{ji}N_j(t) - \sum_{j\neq i} \tilde{B}_{ij}N_i(t)   & \\
      - 2 \beta_{\textrm{LiCs-LiCs}} \overline{n}_{0,\textrm{LiCs}} \frac{\sum_{j}N_j(t)}{N_\textrm{0,\textrm{tot}}} N_i(t) \nonumber . &
  \end{eqnarray}

  As shown in Fig.~\ref{fig:timeevolution3}~b) this leads to a significant faster decay which masks the redistribution process in the integrated vibrational signal while the dynamics remain visible in the single rotational components.

\section{Experimental observation of population dynamics}

  In the experiment ultracold LiCs molecules are formed by photoassociation (PA)~\cite{thorsheim1987,jones2006} of $^{133}$Cs and $^{7}$Li atoms trapped simultaneously in a quasi electrostatic trap (QUEST). After the removal of remaining atoms a pure molecular sample can be kept in the trap for variable times. The QUEST is formed by the single focus of a CO$_2$ laser (power 64(4)~W, waist $w_{0}$=89(4)~$\mu m$) emitting at 10.6~$\mu m$. Details of the QUEST setup can be found in \cite{Staanum2006} and references therein. The resulting trap depths and frequencies for the different species are given in Tab.~\ref{tab:Exp:CO2:depths}. They are calculated from measured values for beam power and waist, experimental values for atomic polarizabilities~\cite{Molof1974a}, and \textit{ab-initio} calculated values for the molecular polarizabilities~\cite{Deiglmayr2008a}.

  \begin{table*}
    \begin{center}
    \resizebox{0.8\textwidth}{!}{
    \begin{tabular}{|r | c | c | c | c | c|}
    \hline
    Species                     & Li        & Cs        & LiCs(\Xstate,$v''$=0) & LiCs(\Xstate,$v''$=42) & LiCs(\astate,$v''$=0) \\ \hline
    Trap depth [$\mu$K]         & 194(21)   & 479(53)   & 435(48)               & 700(77)                & 744(82)  \\
    $\omega_{rad}$/2$\pi$ [1/s] & 1729(166) & 623(60)   & 579(55)               & 735(70)                & 758(73)  \\
    $\omega_{ax}$/2$\pi$ [1/s]  & 47(7)     & 17(2)     & 16(2)                 & 20(3)                  & 20(3)  \\ \hline
    \end{tabular}}
    \vspace{0.5cm}
    \caption{Trap depths and frequencies of the QUEST for the different species. The given error is based on the experimental uncertainties only, no error on the theoretical values was assumed. \Xstate,$v''$=42 experiences the maximum trap depth of all \Xstate levels.}
    \label{tab:Exp:CO2:depths}
    \end{center}
\end{table*}

  Despite its significant detuning, the optical trapping field could drive Raman transitions between neighboring levels due to its high intensity. The rate for such transitions can be estimated from a simple classical model of the Raman process~\cite{Bernath2005}. Using the \textit{ab-initio} calculated polarizability tensor for the \Xstate state of LiCs~\cite{Deiglmayr2008a} and the experimental laser intensity we derive transition rates which are several orders of magnitude below the transition rates for spontaneous decay and BBR transitions. Thus vibrational or rotational heating by the trapping CO$_2$ laser can be neglected even on longer timescales.

  For PA of LiCs molecules in the dipole trap we load first Cs atoms from a Zeeman slowed atomic beam into a conventional MOT setup. After a red-detuned molasses phase 8$\times$10$^5$ Cs atoms are transferred into the QUEST at a density of $1\times10^{11}$~cm$^{-3}$ and a temperature of 33(4)~$\mu$K. Afterwards the Li MOT is loaded also from a Zeeman slowed atomic beam. After magnetic compression typically 2$\times$10$^5$ Li atoms are transferred into the dipole trap at a density of 5$\times10^{9}$~cm$^{-3}$. The comparatively low density has to be attributed to the relatively small initial number of 2$\times$10$^7$ atoms in the MOT, and to a low transfer efficiency due to the small phase-space overlap between MOT and QUEST. Li and Cs atoms are then kept together in the trap for 500\,ms to reduce the Li temperature by thermalization with the Cs ensemble~\cite{Mudrich2002}. The background gas in the science chamber at a pressure of around 2$\times10^{-11}$~mbar leads to atomic lifetimes on the order of 30 seconds.

    \begin{figure}
   \begin{center}
   \includegraphics[width=0.9\columnwidth,clip]{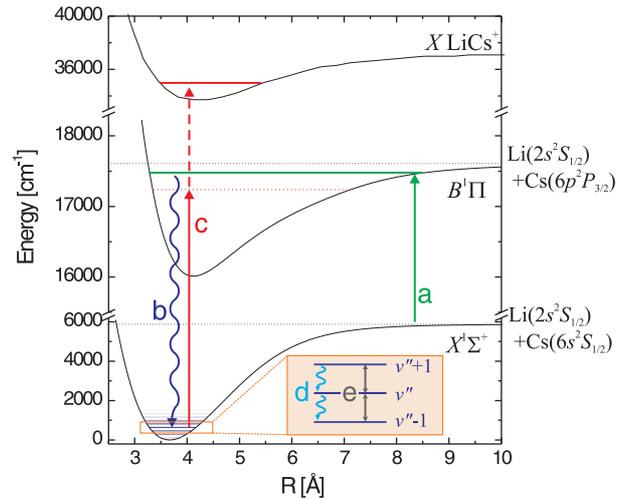}
   \end{center}
   \caption{Exemplary processes in the formation and detection of a ground state LiCs molecule: a) photoassociation from the continuum into the \Bstate state, b) spontaneous decay into the \Xstate, c) two-photon ionization via intermediate levels in the \Bstate state. Redistribution of population: d) spontaneous decay, e) transitions driven by black-body radiation (for clarity only transitions with $\vartriangle$$v'' = \pm 1$ are shown).}
   \label{fig:levelScheme}
  \end{figure}

  After these preparatory steps molecules are formed and detected as illustrated in Fig.~\ref{fig:levelScheme}. First PA is performed by a light pulse (300\,ms long) from a tunable cw TiSa laser (typically 500~mW, focused to 100~$\mu m$). The excited LiCs molecules decay spontaneously within few tens of nanoseconds into different electronic ground state levels. Short pulses of resonant light are used to remove remaining atoms, then all laser beams are blocked by mechanical shutters. After a variable hold time in the dipole trap, molecules in given vibrational ground state levels are ionized by resonant-enhanced multi-photon ionization (REMPI) and the resulting ions are detected in a time-of-flight mass spectrometer. From the number of detected ions we deduce an overall number of roughly 5$\times10^{3}$ molecules in a broad distribution of internal states~\cite{Deiglmayr2009a} which are initially stored in the trap. The translational temperature of the molecules is mostly determined by the temperature of the heavier Cesium and can thus be estimated to be around 50$\mu$K. The density of the molecular ensemble is not easily defined, as the trapping potential depends on the vibrational state. However deeply bound LiCs molecules experience a similar trapping potential as Cs atoms and one can thus derive the molecular density from the measured atomic density as $n_{\textrm{LiCs}}=N_{\textrm{LiCs}}/N_{\textrm{Cs}}\,n_{\textrm{Cs}}$. This yields a peak density of $n_{\textrm{LiCs}}\approx$6$\times10^{8}$~cm$^{-3}$. For molecules in high vibrational levels the density could be up to 60\% higher due to the increased polarizability.

    \begin{figure}
   \begin{center}
   \includegraphics[width=0.85\columnwidth,clip]{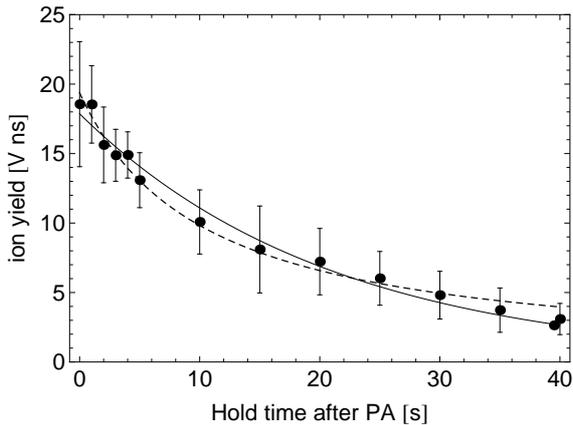}
   \caption{Ion yield at 16893.3\rcm and high intensity ($\sim$350 J/cm$^2$) after a varying hold time in the QUEST. Under these REMPI conditions molecules in levels \Xstate,$v''$$\scriptstyle\gtrsim$30 can be detected. Every point is an average of nine measurements, shown uncertainties are the standard deviation of the measurement. The solid (dashed) line shows a fit of a first-order (second-order) decay process to the data.}
   \label{fig:timeevolutionHigh}
   \end{center}
  \end{figure}

  The slowest decay from the trap is observed for molecules in high lying levels, shown in Fig.~\ref{fig:timeevolutionHigh}. Assuming a first order decay process a fit to the data yields a time constant of 21(1) seconds (comparable to the lifetime of pure atomic samples in the trap). At our estimated molecular density bi-molecular collisions might also become significant. Fitting the decay curve with a pure two-body decay yields a rate coefficient of $\beta_{\textrm{LiCs-LiCs}}$=2$\times10^{-10}$~cm$^{3}/s$ for inelastic LiCs-LiCs collisions (making the above discussed assumptions about the molecular density). This is a realistic value when compared with rate coefficients found in other experiments. For collisions of KRb molecules in deeply bound, distinguishable states a rate coefficient of $2\times10^{-10}$~cm$^{3}/s$ ~\cite{Ospelkaus2010} was measured. For RbCs molecules in different weakly bound triplet states a value on the order of 1$\times10^{-10}$~cm$^{3}/s$ was found~\cite{Hudson2008}. However our data doesn't allow us to distinguish between a first or second order decay process. As the observed loss rate is comparable to the atomic ones, we cannot exclude that the observed molecular trap loss is due to collisions with background gas or another yet unidentified loss mechanism.

  At typical REMPI intensities of 2~J/cm$^2$ the observed REMPI resonances have a width of 2-3~\rcm (full width at half maximum), which is very likely due to power broadening of the REMPI transitions (the line width of the ionization laser is 0.07~\rcm, the spread of a typical initial rotational distribution $\sim$0.5~\rcm). Simulations of the REMPI line profile after different storage times (and thus different rotational distributions) show that the ionization signal at the initial peak position of the REMPI signal is a reliable measurement for the total population in a vibrational level.

    \begin{figure}[htb]
   \begin{center}
   \includegraphics[width=0.85\columnwidth,clip]{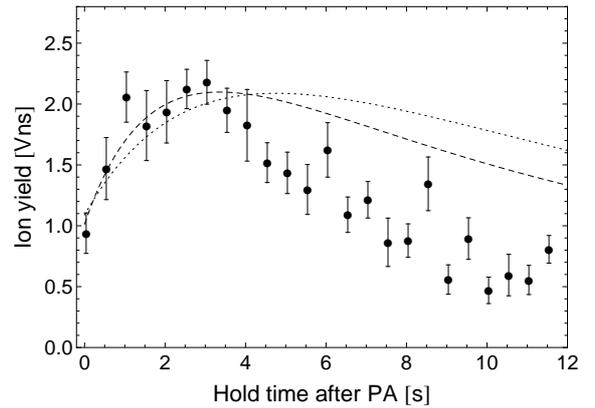}
   \caption{Ion yield from $v''$=3 molecules ionized at 16886.3\rcm and low intensity ($\sim$2 J/cm$^2$)after a varying hold time in the QUEST. Every point is an average of 15 measurements, given uncertainties are the standard deviation of the mean value. The dotted (dashed) line shows the simulated signal for the adjusted first (second) order model as discussed in the text.}
   \label{fig:timeevolution3exp}
   \end{center}
  \end{figure}

   Fig.~\ref{fig:timeevolution3exp} shows the ion yield from molecules in the vibrational level \Xstate,$v''$=3. We used depletion spectroscopy to check that at the employed REMPI energy of 16886.3~\rcm the detected ions originate indeed predominantly from this ground state level. The experimental curve shows the same qualitative behavior as seen for the simulation assuming a first order loss process (Eq.~\ref{eq:rateeq}): first an increase is observed, then a maximum is reached after around 3s followed by a decay of the signal. However the modulation of the experimental curve is significantly stronger than in the simulation. The fast decay for hold times longer than 3s is in fact better described by the model assuming a second order loss process. In order to adjust the model to the data we vary only the relative initial population in the vibrational levels $v''$=3 and $v''$=4 which determines most strongly the modulation of the population in $v''$=3. The initial population is the most sensitive quantity in the model as it is mostly determined by the Franck-Condon (FC) factors between the PA level and the electronic ground state levels which are sensitive to small deviations of the potential energy curves. The subsequent dynamic of the population in the ground state on the other hand is dominated by transitions between neighboring levels in the \Xstate state for which the rates are less sensitive to the exact shape of the potential. Fig.~\ref{fig:timeevolution3exp} shows the result of simulations where the ratio $N_{v"=4}$/$N_{v"=3}$ was adjusted to reproduce the experimental modulation strength. For the first order model of Eq.~\ref{eq:rateeq} the ratio was increased by a factor of 2.7, for the second order model of Eq.~\ref{eq:rateeq2b} by a factor of 5.0. The latter simulation reproduces the initial modulation of the experimental data very well. However the necessary adjustment of the starting conditions is stronger than one would expect for FC factor calculation from accurate experimental potentials. The experimentally observed decay of the population in $v''$=3 is still faster than expected from our measurements of the decay for high lying vibrational levels. This could be an indication for an increase of the cross section for inelastic LiCs-LiCs collisions with increasing binding energies as predicted for the case of RbCs~\cite{Kotochigova2010}.

\section{Conclusion and outlook}

   We have observed population redistribution caused by BBR and spontaneous decay for ultracold LiCs molecules stored in an optical dipole trap. Similar rate calculations for other polar molecules which have been formed in the ultracold regime, KRb~\cite{Ni2008} and RbCs~\cite{Sage2005}, show that the timescales for redistribution processes in these systems is orders of magnitude larger~\cite{Vanhaecke2007,Buhmann2008}. Due to the large electric dipole moment of ground state LiCs molecules and the associated strong variation of the dipole moment function with the internuclear distance, radiative and BBR-driven dipole transitions occur in this system on a timescale relevant for this and future experiments. In the specific case of LiCs, molecules prepared in the rovibrational ground state will be pumped out of this level by BBR-driven transitions with a time constant of 65~s. If in future experiments this imposes a limitation, this value could be significantly lengthened by cooling the apparatus. The comparison of our experimental results with models based on \textit{ab initio} calculated rates may indicate the presence of a bimolecular loss with a rate coefficient exceeding 2$\times10^{-10}$~cm$^{3}/s$ for deeply bound LiCs molecules.

\begin{acknowledgement}
   This work is supported by the DFG under WE2661/6-1 in the framework of the Collaborative Research Project QuDipMol within the ESF EUROCORES EuroQUAM program. JD acknowledges partial support of the French-German University.
\end{acknowledgement}
%\bibliography{mixtures}
%\input{deiglmayr.bbl}

\end{document}